\documentclass[letterpaper, 10 pt, conference]{ieeeconf}  

\IEEEoverridecommandlockouts                              
\usepackage{graphicx} 
\usepackage{amsmath} 
\usepackage{amssymb}  
\usepackage{booktabs}
\usepackage{multirow}
\usepackage{algorithm}
\usepackage{algpseudocode}
\usepackage{xcolor}
\usepackage{hyperref}
\hypersetup{
  colorlinks=true,
  linkcolor=red,
  citecolor=blue,
  urlcolor=blue
}

\makeatletter
\def\ps@copyright{%
  \let\@mkboth\@gobbletwo
  \def\@oddhead{}%
  \let\@evenhead\@oddhead
  \def\@oddfoot{%
    \raisebox{-8pt}[0pt][0pt]{%
      \parbox{\textwidth}{%
        \centering\normalfont\fontsize{10}{10}\selectfont
        This work has been submitted to the IEEE for possible publication. Copyright may be transferred without notice, after which this version may no longer be accessible.}}}%
  \let\@evenfoot\@oddfoot
}
\makeatother

\title{\LARGE \bf
Distributed ToA Localization of Acoustic Sources with Unknown Time of Emission via Operator Splitting
}

\author{Anton Tolstonogov, David Cabecinhas, Pedro Batista, and Antonio Pascoal
\thanks{This work was supported 
        by FCT funding PRT/BD/155064/2023 (DOI: 10.54499/PRT/BD/155064/2023),
        LARSyS FCT funding (DOI: 10.54499/LA/P/0083/2020, 10.54499/UIDP/50009/2020, and 10.54499/UIDB/50009/2020),
        and the FCT India-Portugal project RAIECO (ID: DRI/India/0699/2020).}
\thanks{
    Anton Tolstonogov, David Cabecinhas, Pedro Batista, and Antonio Pascoal are with
    Institute for Systems and Robotics (ISR), LARSyS, Instituto Superior Técnico (IST), University of Lisbon, Portugal,
    {\tt\small e-mail: anton.tolstonogov@tecnico.ulisboa.pt, dcabecinhas@isr.tecnico.ulisboa.pt, pbatista@isr.tecnico.ulisboa.pt, antonio.pascoal@tecnico.ulisboa.pt}}%
}

\makeatletter
\algnewcommand{\LineComment}[1]{%
  \Statex \hskip\ALG@thistlm {\footnotesize\color{gray}\(\triangleright\)\ \textit{#1}}%
}
\makeatother
\makeatletter
\algnewcommand{\LineCommentFirst}[1]{%
  \Statex \hskip\dimexpr\ALG@thistlm+\algorithmicindent\relax
  {\footnotesize\color{gray}\(\triangleright\)\ \textit{#1}}%
}
\makeatother

\algrenewcommand\algorithmiccomment[1]{%
  \unskip\hspace{0.5em}{\footnotesize\color{gray}\(\triangleright\)\ \textit{#1}}%
}

\begin{document}

\maketitle
\thispagestyle{copyright}
\pagestyle{empty}

\begin{abstract}
Localization of non-cooperative acoustic sources using multiple spatially distributed receivers is critical for applications such as marine-life tracking, search-and-rescue operations, and maritime security in underwater environments.
In conventional Time of Arrival (ToA) systems, the emission time is typically known explicitly or implicitly through clock synchronization or two-way communication, so the problem reduces to range-based localization.
In passive settings, however, only reception timestamps are available.
Thus, the emission time must be eliminated through Time Difference of Arrival (TDoA) preprocessing or estimated jointly with the source position.
For the case of a single source, we address the distributed localization problem over a receiver network by reformulating passive localization directly as a ToA problem with unknown time of signal emission.
This yields a distributed consensus optimization problem, which we solve using an operator-splitting method, namely an edge-based Distributed Alternating Direction Method of Multipliers (DADMM) scheme that decomposes the estimation task into local subproblems coupled through agreement constraints.
We derive closed-form local update equations for the local DADMM subproblems and establish convergence properties for a smoothed approximation of the measurement model.
Numerical simulations illustrate the efficacy of the proposed approach.
\end{abstract}

\section{INTRODUCTION}

The surge of interest worldwide in ocean-related activities is steadily driving the development of innovative technologies to monitor and track
underwater objects and biological entities in open and often unstructured environments.
In particular, applications such as marine-life tracking \cite{haysTranslatingMarineAnimal2019, espinozaTestingNewAcoustic2011}, search-and-rescue missions \cite{OperationalSearchMH370}, and maritime security \cite{gebbiePassiveAcousticArray2011, hamiltonAntisubmarineWarfareApplications2010} frequently require localization of non-cooperative underwater acoustic sources.
Since these scenarios involve operations in inherently GNSS-denied conditions and without transmitter--receiver synchronization, accurate localization remains a fundamental challenge.
In the setting considered here, a non-cooperative acoustic source is observed by multiple spatially distributed receivers with known positions that cooperate over a communication network to estimate the source location.

Acoustic source localization relies on several fundamental measurement modalities, including received signal strength (RSS), angle of arrival (AoA), time of arrival (ToA), time difference of arrival (TDoA), and frequency difference of arrival (FDoA) \cite{bhardwajReviewLocalizationAlgorithm2024, yangComprehensiveSurveyUnderwater2025}.
These modalities differ in their physical observables: RSS is based on attenuation models, AoA on directional sensing, and timing-based approaches (ToA, TDoA) on propagation delays.

Localization techniques can also be classified as active or passive, depending on how the localization system interacts with a source \cite{yangComprehensiveSurveyUnderwater2025}.
Active methods require coordinated signaling or two-way ranging, whereas passive methods rely on one-way emissions from an unsynchronized source.
In many ToA systems, the emission time is available explicitly or implicitly through synchronized transmissions, which makes the measurements equivalent to ranges.
In the non-cooperative passive scenarios considered here, however, only reception timestamps are available at the receivers, while the emission time remains unknown.
This leaves two main options: eliminate the emission time through TDoA preprocessing or estimate it jointly with the target position in a ToA formulation with unknown time of signal emission.

TDoA methods eliminate the unknown time of signal emission by forming pairwise differences between receivers' arrival-time measurements.
This transformation introduces stronger coupling between receivers and increases the structural complexity of the estimation problem, as each measurement is inherently associated with a pair of nodes.

Despite the above, the TDoA principle remains attractive in practice, since it can provide high positioning accuracy \cite{wangTDOASourceLocalization2013} while requiring lower hardware complexity, reduced size and power consumption, and simplified deployment compared to AoA systems \cite{ComparisonTimedifferenceofarrival2018}.
These properties make TDoA particularly suitable for large-scale passive acoustic networks operating in constrained underwater environments.

Centralized methods dominate TDoA-based localization, with all measurements processed at a fusion node.
Within this class, geometric and algebraic solvers remain widely used: robust weighted-least-squares formulations with geometric constraints improve stability in hyperbolic positioning \cite{jinRobustTDOALocalization2018}, and algebraic solutions jointly exploit TDoA/FDoA measurements to recover both position and velocity of a target~\cite{hoAccurateAlgebraicSolution2004}.
Alongside these, maximum-likelihood approaches provide statistically principled estimators, ranging from direct position determination for stationary emitters \cite{vankayalapatiTDOA2014} to machine learning formulations capable of operating directly on compressed measurements \cite{caoMaximumLikelihood2017}.
More recently, convex optimization techniques have been introduced to improve robustness and provide stronger convergence guarantees, including semidefinite programming (SDP) and second-order cone programming (SOCP) relaxations that convexify the underlying nonconvex hyperbolic constraints \cite{yangOptimalSensorPlacement2024, sathyanTwoSolutions2007}.

Unlike ToA- or AoA-based models, TDoA measurements are inherently pairwise.
Consequently, no receiver can evaluate its contribution using its own data only.
Although the source state is globally shared, the data required to evaluate each residual is distributed across receiver pairs.
This gap between measurement structure and communication topology limits the applicability of standard distributed estimation schemes, such as sequential fusion, average consensus, gossip, and diffusion methods \cite{heDistributedEstimationLowcost2020}, because they assume that each agent can compute a well-defined local estimate from locally available data and iteratively reconcile it with its neighbors.
In TDoA settings, however, no individual receiver has sufficient information to do so.

To overcome this structural limitation, we keep the problem in the ToA domain and treat the emission time as an additional unknown, so that each receiver is associated with its own locally available ToA measurement.
Although a single ToA measurement cannot uniquely determine both the source position and the emission time, the local models together define a global optimization problem that can be solved cooperatively over the receiver network.
We solve the reformulated problem using a Distributed Alternating Direction Method of Multipliers (DADMM) \cite{schizasConsensusAdHoc2008} scheme that decomposes the objective into local subproblems and enforces agreement on shared variables through consensus constraints and dual updates exchanged between neighboring agents.

In contrast to prior works that employ ADMM primarily as a computational optimization tool within a centralized framework \cite{zhuADMMBasedTDOAEstimation2018,tausiesakulTDoALocalizationWireless2025, liImprovedTwoStepConstrained2020}, we use ADMM as a mechanism for distributing the estimation process itself across networked agents, enabling a cooperative solution of the global problem without central aggregation.

The main contributions of this work are as follows.
\begin{itemize}
\item We reformulate passive localization as a ToA problem with unknown time of signal emission in a way that is compatible with distributed optimization over a receiver network.
\item We derive an efficient DADMM-based distributed method for solving the reformulated problem, with a closed-form local update and without nested numerical optimization at each iteration.
\item We establish convergence properties of the smoothed approximation of the measurement model and identify a sufficient range of penalty parameters for the DADMM approach.
\end{itemize}

The remainder of this paper is organized as follows.
Section~\ref{sec:problem_statement} introduces the problem formulation, including the receiver-network description and the measurement function.
Section~\ref{sec:localization_centralized} casts the localization problem with unknown time of signal emission as an optimization problem in a centralized scenario.
Section~\ref{sec:localization_decentralized} reformulates the localization problem in a distributed setting via the DADMM approach and provides closed-form solutions for each iteration in the procedure.
Section~\ref{sec:convergence_analysis} provides convergence properties of the approach.
Section~\ref{sec:numerical_simulation} reports numerical results and comparisons with a centralized nonlinear solver and a CRLB reference.
Section~\ref{sec:conclusion} concludes the paper and outlines future work.

\section{PROBLEM STATEMENT}
\label{sec:problem_statement}
For clarity of exposition, we consider a single target.
If multiple targets are distinguishable, for example by frequency content or other signal features, the proposed formulation applies independently to each target, and the overall problem decouples into separate single-target estimation problems.

\subsection{Network Model}
\label{ssec:network_model}
The network of receivers considered throughout this paper is represented as an undirected, connected graph $\mathcal{G} = (\mathcal{N}, \mathcal{E})$.
The node set $\mathcal{N} = \{1,2,\ldots, N\}$ consists of $N$ nodes (receivers).
For each node $i \in \mathcal{N}$, we define $\mathcal{N}_i = \{j \mid (i,j) \in \mathcal{E}\}$ as the set of its adjacent nodes and $N_i$ as the cardinality of $\mathcal{N}_i$.
The neighborhood topology is assumed to be known at each node $i$.
The position of receiver $i$ in a common reference frame is represented by $\mathbf{s}_i \in \mathbb{R}^M$ for all $i \in \mathcal{N}$.

\subsection{Measurement Model}
The noisy ToA measurement collected at receiver $i$ from the target is denoted by $\tau_{i}$.
This measurement can be expressed as
\begin{equation}
\label{eq:noise}
\tau_{i} = t + \frac{1}{v}\|\mathbf{p} - \mathbf{s}_{i}\| + w_i,
\end{equation}
where $t$ is the (unknown) signal emission time, $v$ is the known acoustic propagation velocity, $\mathbf{p} \in \mathbb{R}^M$ is the unknown target position, $M \in \{2,3\}$ is the spatial dimension (2D or 3D), and $w_i \sim \mathcal{N}(0, \sigma^2)$ is zero-mean independent and identically distributed Gaussian noise with variance $\sigma^2$.

\begin{figure}[thpb]
    \centering
    \includegraphics[width=7.0cm]{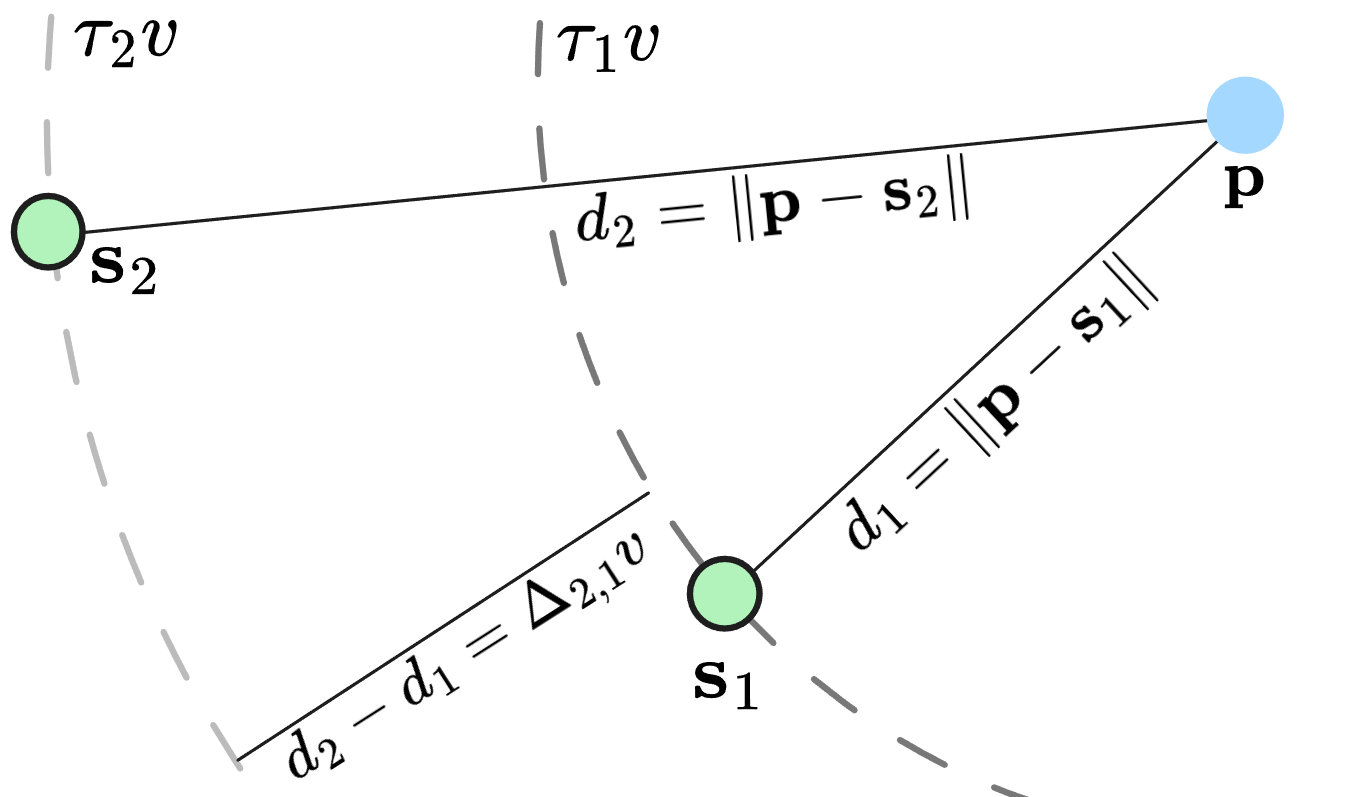}
    \caption{Measurement geometry for the case of two receivers and one target ($p$ denotes the emitting target, and $s_i$, $i=1,2$, denote the receivers).\label{fig:geometry}}
\end{figure}

\subsection{Objective}
Assume that the positions of at least $N \ge 3$ receivers for $M=2$ or $N \ge 4$ for $M=3$ are known and that their clocks are synchronized with sufficient accuracy (e.g., GNSS-referenced timing for surface platforms, or pre-mission synchronization for underwater platforms).
Each receiver measures a noisy signal arrival time $\tau_i$ for $i \in \mathcal{N}$ from a target whose emission time is unknown. Under these assumptions, our objective is to develop a fully decentralized method for estimating the target position based solely on the time-of-arrival measurements.

\section{LOCALIZATION PROBLEM FORMULATION}
\label{sec:localization_centralized}
Based on the measurement model~\eqref{eq:noise}, the corresponding least-squares problem can be formulated as
\begin{equation}
    \label{eq:opt_problem}
    \min_{\mathbf{p}, t}
    \sum_{i=1}^{N}
    \frac{1}{2}
    (\tau_i - t - \frac{1}{v}\|\mathbf{p} - \mathbf{s}_{i}\|)^2.
\end{equation}

In a traditional centralized approach (or in a decentralized approach with a fusion center), a central node either applies a nonconvex solver directly or reformulates the problem in a TDoA setting to eliminate the variable $t$.
In the latter case, the joint estimation problem is formulated by considering pairwise differences of the arrival times among sensor nodes.
The time-difference-of-arrival measurements can be obtained through the preprocessing step between receivers $i$ and $j$ given by
\begin{equation}
    \label{eq:hyperbolic}
    v\Delta_{i,j} = v(\tau_i - \tau_j) = \|\mathbf{p} - \mathbf{s}_i\| - \|\mathbf{p} - \mathbf{s}_j\| + w_i - w_j,
\end{equation}
where $\Delta_{i,j}$ is the TDoA measurement and $v\Delta_{i,j}$ is the range difference between two receivers relative to the target position (see Figure~\ref{fig:geometry}).

However, there are several drawbacks to this approach~\cite{xuSourceLocalizationWireless2011}:
\,(i)~the noise terms in pairs of TDoA measurements are no longer independent, and measurements $\Delta_{i,j}$ and $\Delta_{k,j}$ are correlated since they share a common reference;
\,(ii)~in a distributed setting, TDoA requires an additional initial communication step to form the pairwise differences;
\,(iii)~the TDoA measurement in~\eqref{eq:hyperbolic} yields a hyperbolic observation model, which is challenging to solve.

In this work, we show that~\eqref{eq:opt_problem} provides a convenient formulation for distributed optimization that avoids explicit TDoA preprocessing and admits closed-form local updates within the proposed iterative scheme.

\section{DISTRIBUTED LOCALIZATION VIA ADMM}
\label{sec:localization_decentralized}
Following the edge-based DADMM formulation in~\cite{chagantiDistributedADMMTarget2025, maHybridADMMUnifying2018}, we rewrite~\eqref{eq:opt_problem} in a distributed form.
For each receiver $i \in \mathcal{N}$, we define a local state variable $\mathbf{x}_i = [\mathbf{p}_i, t_i]^\top$, where $\mathbf{p}_i$ is the local estimate of the target position at receiver $i$ and $t_i$ is the local estimate of the emission time.
To impose agreement across neighboring receivers, we further introduce an edge variable $\mathbf{y}_{i,j}$ for each pair of connected receivers $i$ and $j$.
The resulting distributed optimization problem is
\begin{equation}
    \label{eq:optimization_problem_local}
    \begin{aligned}
    \min_{\{\mathbf{x}_i\}, \{\mathbf{y}_{i,j}\}}
    \quad & \sum_{i \in \mathcal{N}} f_i(\mathbf{x}_i) \\
    \text{s.t.}\quad &
    \mathbf{x}_i = \mathbf{y}_{i,j},\;
    \mathbf{x}_j = \mathbf{y}_{i,j},
    \quad \forall (i,j) \in \mathcal{E},
    \end{aligned}
\end{equation}
where $\{\mathbf{x}_i\}_{i \in \mathcal{N}}$ denotes the collection of local state variables, $\{\mathbf{y}_{i,j}\}_{(i,j) \in \mathcal{E}}$ denotes the collection of edge consensus variables, $\mathcal{E}$ denotes the set of edges introduced in Section~\ref{ssec:network_model}, and
\[
f_i(\mathbf{x}_i) = \frac{1}{2} ( \tau_{i} - t_i - \frac{1}{v}\|\mathbf{p}_i - \mathbf{s}_{i}\| )^2
\]
is the local cost function at receiver $i$.

\begin{figure}[thpb]
    \centering
    \includegraphics[width=8.0cm]{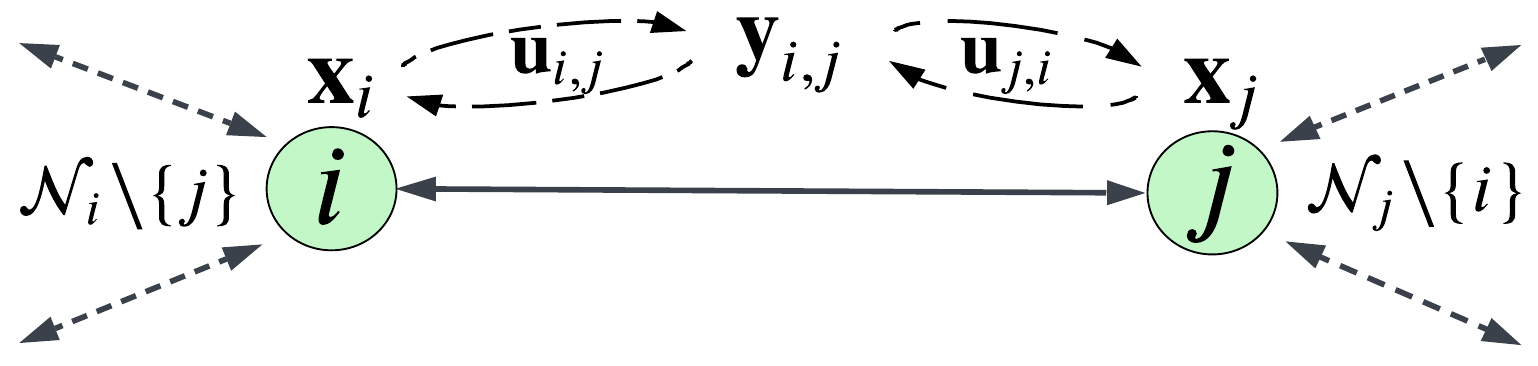}
    \caption{The edge variable $\mathbf{y}_{i,j}$ enforces agreement between the states $\mathbf{x}_i$ and $\mathbf{x}_j$, and each constraint is associated with a corresponding dual variable $\mathbf{u}_{i,j}$ or $\mathbf{u}_{j,i}$.\label{fig:dadmm_variables}}
\end{figure}

For each constraint $\mathbf{x}_i - \mathbf{y}_{i,j} = 0$ and $\mathbf{x}_j - \mathbf{y}_{i,j} = 0$ in~\eqref{eq:optimization_problem_local}, we introduce the corresponding Lagrangian dual variables $\mathbf{u}_{i,j}$ and $\mathbf{u}_{j,i}$, as illustrated in Figure~\ref{fig:dadmm_variables}.

Hence, the augmented Lagrangian $\mathcal{L}_{i-j}$ for the edge connecting receivers $i$ and $j$ takes the scaled form
\begin{equation}
    \begin{aligned}
        \mathcal{L}_{i-j} &= 
        \frac{1}{2}\|\mathbf{x}_i - \mathbf{y}_{i,j} + \mathbf{u}_{i,j}\|_\Theta^2
        - \frac{1}{2}\|\mathbf{u}_{i,j}\|_\Theta^2 + \\
        &+ \frac{1}{2}\|\mathbf{x}_j - \mathbf{y}_{i,j} + \mathbf{u}_{j,i}\|_\Theta^2 
        - \frac{1}{2}\|\mathbf{u}_{j,i}\|_\Theta^2,
    \end{aligned}
\end{equation}
where
\begin{equation}
    \label{eq:theta}
    \Theta =
    \begin{bmatrix}
    \rho_p I_M & 0 \\
    0 & \rho_t
    \end{bmatrix}
\end{equation}
weights the position and emission-time components with different penalty parameters, accounting for their distinct scales and sensitivities, and $I_M \in \mathbb{R}^{M\times M}$ is an identity matrix.

The local Lagrangian $\mathcal{L}_i$ associated with the function $f_i(\mathbf{x}_i)$, obtained by collecting the terms associated with receiver $i$ over all neighboring edges $j \in \mathcal{N}_i$, becomes
\begin{equation}
    \mathcal{L}_{i} = f_i(\mathbf{x}_i) + 
    \sum_{j \in \mathcal{N}_i}
    (
        \frac{1}{2}\|\mathbf{x}_i - \mathbf{y}_{i,j} + \mathbf{u}_{i,j}\|_\Theta^2
        - \frac{1}{2}\|\mathbf{u}_{i,j}\|_\Theta^2
    ).
\end{equation} 

Then, under the DADMM framework, each agent $i$ iteratively updates its estimate of the target state according to
\begin{equation}
    \label{eq:prox_x}
    \mathbf{x}_i^{[k+1]} = \arg \min_{\mathbf{x}_i}
        f_i(\mathbf{x}_i) + \frac{1}{2}\sum_{j \in \mathcal{N}_i}
        \|(\mathbf{x}_i - \mathbf{y}_{i,j}^{[k]} + \mathbf{u}_{i,j}^{[k]})\|_\Theta^2
    ,
\end{equation}
followed by the edge-variable update
\begin{equation}
    \label{eq:prox_y}
    \mathbf{y}_{i,j}^{[k+1]} = \frac{1}{2}
    \left(
        \mathbf{x}_i^{[k+1]} + \mathbf{x}_j^{[k+1]} + \mathbf{u}_{i,j}^{[k]} + \mathbf{u}_{j,i}^{[k]}
    \right),
\end{equation}
and by the local dual update $\mathbf{u}_{i,j}^{[k+1]}$
\begin{equation}
    \label{eq:prox_u}
    \mathbf{u}_{i,j}^{[k+1]} =
    \mathbf{u}_{i,j}^{[k]} + 
    \mathbf{x}_i^{[k+1]} -
    \mathbf{y}_{i,j}^{[k+1]}.
\end{equation}

To avoid solving~\eqref{eq:prox_x} numerically at each iteration, we reparameterize the local position as $\mathbf{p}_i = \mathbf{s}_i + r_i \mathbf{n}_i$, where $r_i \ge 0$ and $\|\mathbf{n}_i\| = 1$.
Under this parameterization, the local minimizer of~\eqref{eq:prox_x} can be obtained in closed-form from the current local data and consensus variables.
The derivation is given in Appendix~\ref{sec:app_proximal_x}.

Since the objective in~\eqref{eq:optimization_problem_local} is nonconvex, the attained solution depends on the initialization.
Hence, in the absence of prior information, we adopt a cold-start initialization at $k=0$, $\mathbf{x}_i^{[0]} = [\mathbf{p}_i^{[0]}, t_i^{[0]}]^\top$, constructed from locally available geometric information as
\begin{equation}
\label{eq:init_x}
\mathbf{p}_i^{[0]} =
\frac{1}{N_i + 1}
(
\mathbf{s}_i + \sum_{j \in \mathcal{N}_i}\mathbf{s}_j
),
\quad
t_i^{[0]} = \tau_i - \frac{1}{v}\|\mathbf{p}_i^{[0]} - \mathbf{s}_i\|,
\end{equation}
which sets the initial position to the local receiver centroid and makes the local residual in $f_i$ equal to zero.
In the degenerate case where this centroid coincides with $\mathbf{s}_i$, any sufficiently small offset satisfying $\|\mathbf{p}_i^{[0]} - \mathbf{s}_i\| > 0$ can be used instead.
For subsequent measurements, the previous estimate can be reused as a warm start, thereby reducing the number of ADMM iterations required for convergence.

\begin{figure}[thpb]
    \centering
    \includegraphics[width=8.0cm]{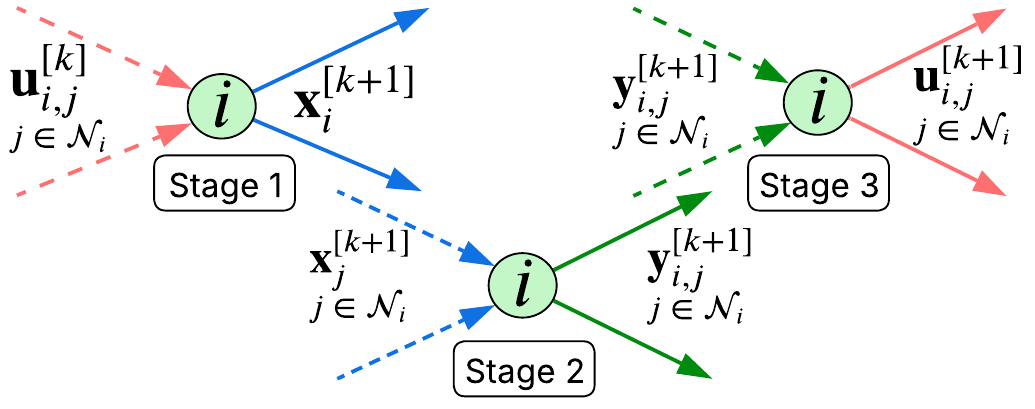}
    \caption{Data flow of the distributed localization procedure according to Algorithm~\ref{alg:dlm} for receiver $i$.\label{fig:data_flow}}
\end{figure}

The complete distributed localization procedure, including the initialization, local subproblem updates, dual-variable updates, and stopping criteria, is summarized in Algorithm~\ref{alg:dlm} and illustrated in Figure~\ref{fig:data_flow}.

\begin{algorithm}[t]
\caption{Distributed algorithm for ToA-based localization with an unknown signal emission time\label{alg:dlm}}
\begin{algorithmic}[1]
\Require
Graph $\mathcal{G} = (\mathcal{N}, \mathcal{E})$, $\sigma$, $\rho_t$, $\rho_p$, $\varepsilon_\text{conv}$, and $\varepsilon_\text{feas}$.

Additionally, $\mathcal{N}_i = \{j \mid (i,j) \in \mathcal{E}\}$, $\mathbf{s}_i$, and $\tau_i$ for $i \in \mathcal{N}$.

\For{$i \in \mathcal{N}$} \Comment{Preparation, in parallel}
    \State Initialize $\mathbf{x}_i^{[0]}$ according to \eqref{eq:init_x}
    \State Send $\mathbf{x}_i^{[0]}$ to all neighbors $j \in \mathcal{N}_i$
    \For{$j \in \mathcal{N}_i$}
        \State Initialize $\mathbf{y}_{i,j}^{[0]} = \frac{1}{2}\left(\mathbf{x}_i^{[0]} + \mathbf{x}_j^{[0]}\right)$
        \State Initialize $\mathbf{u}_{i,j}^{[0]} = 0$
    \EndFor
\EndFor
\For{$k=0, 1, 2, \ldots$} \Comment{DADMM sequence}
    \LineCommentFirst{For each receiver $i$ in parallel}
    \LineCommentFirst{Stage 1}
    \State Send $\mathbf{u}_{i,j}^{[k]}$ to all neighbors $j \in \mathcal{N}_i$
    \State Update $\mathbf{x}_i^{[k+1]}$ (see Appendix~\ref{sec:app_proximal_x})
    \State Send $\mathbf{x}_i^{[k+1]}$ to all neighbors $j \in \mathcal{N}_i$
    \LineComment{Stage 2}
    \State Update $\mathbf{y}_{i,j}^{[k+1]}$ according to \eqref{eq:prox_y}
    \State Send $\mathbf{y}_{i,j}^{[k+1]}$ to all neighbors $j \in \mathcal{N}_i$
    \LineComment{Stage 3}
    \State Update $\mathbf{u}_{i,j}^{[k+1]}$ according to \eqref{eq:prox_u}
    \LineComment{Termination check}
    \State Calculate $r_i = \max_{j\in\mathcal{N}_i} \| (\mathbf{x}_i^{[k+1]} - \mathbf{y}_{i,j}^{[k+1]})\|_\Theta$
    \State Calculate $s_i = |\mathcal{N}_i|\|(\mathbf{x}_i^{[k+1]} - \mathbf{x}_i^{[k]})\|_\Theta$
    \If{$r_i \le \varepsilon_\text{feas}$ and $s_i \le \varepsilon_\text{conv}$}
        \State Terminate the iterations at receiver $i$.
    \EndIf
\EndFor
\end{algorithmic}
\end{algorithm}

\section{CONVERGENCE ANALYSIS}
\label{sec:convergence_analysis}
The original objective in \eqref{eq:optimization_problem_local} is nonconvex and nondifferentiable at $\mathbf{p}_i = \mathbf{s}_{i}$ (for any $i \in \mathcal{N}$) due to the norm operator.

\noindent In the convergence analysis, we use a smoothed approximation given by
\begin{equation}
    \label{eq:r_relaxation}
    \|\mathbf{p}_i - \mathbf{s}_{i}\| \approx \sqrt{\|\mathbf{p}_i - \mathbf{s}_{i}\|^2 + \varepsilon} := r_i (\mathbf{p}_i),
\end{equation}
where $\varepsilon > 0$ is a small relaxation parameter.
This relaxation smooths the behavior of the measurement model near the receiver position $\mathbf{s}_i$.
Interpreting $\varepsilon \approx R_\text{min}^2$, the condition $\sqrt{\varepsilon} \ll \|\mathbf{p}_i - \mathbf{s}_{i}\|$ corresponds to a minimum radius $R_\text{min}$ around $\mathbf{s}_{i}$ within which we do not expect the target to appear, such that
\begin{equation}
    \label{eq:e_meaning}
    \varepsilon \approx R_\text{min}^2.
\end{equation}
In practice, this $R_\text{min}$ distance is negligible relative to the estimated distance to the target.
With this approximation, our smoothed objective becomes
\begin{equation}
    \label{eq:optimization_function_smoothed}
    \sum_{i=1}^{N}
    f_i(\mathbf{p}_i, t_i) =
    \sum_{i=1}^{N}
    \frac{1}{2} \left( \tau_{i} - t_i - \tfrac{1}{v}r_i (\mathbf{p}_i) \right)^2.
\end{equation}

Each function $f_i(\mathbf{p}_i, t_i)$, $i \in \mathcal{N}$, is Lipschitz-differentiable on the analysis region
\begin{equation}
    \Omega_i =
    \left\{(\mathbf{p}_i, t_i):
    r_i (\mathbf{p}_i) \ge R_\text{min},\;
    \left|\tau_i - t_i - \tfrac{1}{v}r_i(\mathbf{p}_i)\right| \le k\sigma
    \right\},
\end{equation}
where $k$ is a user-specified confidence multiplier that defines a trust region scaled by the ToA measurement noise standard deviation $\sigma$ in~\eqref{eq:noise}.
The Lipschitz constants of the subfunctions $f_i(\mathbf{p}_i, t_i)$ are identical across agents and equal to
\begin{equation}
    L_p^i =
    \Big( \frac{1}{v^2} + \frac{k \sigma}{v R_\text{min}}\Big) 
    \text{ and }
    L_t^i = 1
\end{equation}
for the target state and the emission time, respectively.
See Appendix~\ref{sec:app_lipschitz} for details.
Hence, the global block-wise Lipschitz bounds for the smoothed optimization problem~\eqref{eq:optimization_function_smoothed} are
\begin{equation}
    L_p =
    |\mathcal{N}| \Big( \frac{1}{v^2} + \frac{k \sigma}{v R_\text{min}}\Big) 
    \text{ and }
    L_t = |\mathcal{N}|,
\end{equation}
based on the fact that $\|\nabla^2 f(x)\| \le \sum_{i \in \mathcal{N}} \|\nabla^2f_i(x)\|$.

Due to the smoothing in~\eqref{eq:r_relaxation}, the objective in~\eqref{eq:optimization_function_smoothed} is continuously differentiable on $\Omega := \prod_{i=1}^{N}\Omega_i$, where $\Omega$ is the corresponding analysis region. The global constants $L_p$ and $L_t$ therefore provide block-wise Lipschitz bounds for the position and emission-time components of the smoothed objective on $\Omega$.
Let $\varphi_i(\mathbf{x}_i;\rho_p,\rho_t)$ denote the objective of the local smoothed counterpart of~\eqref{eq:prox_x}, and let $\gamma_i(\rho_p,\rho_t)$ denote its strong convexity modulus on $\Omega_i$, that is,
$\gamma_i(\rho_p,\rho_t) = \inf_{\mathbf{x}_i \in \Omega_i} \lambda_{\min}\!\left(\nabla_{\mathbf{x}_i}^2 \varphi_i(\mathbf{x}_i;\rho_p,\rho_t)\right)$.
Define $\gamma_{\min}(\rho_p,\rho_t) := \min_{i \in \mathcal{N}} \gamma_i(\rho_p,\rho_t)$.
Following the convergence framework of~\cite{hongConvergenceAnalysis2016}, we assume that the smoothed objective is bounded below and that
\begin{equation}
    \label{eq:penalty_rules}
    \begin{aligned}
    \rho_p &\ge L_p,\qquad
    \rho_t \ge L_t,\\
    \rho_p\,\gamma_{\min}(\rho_p,\rho_t) &> 2L_p^2,\qquad
    \rho_t\,\gamma_{\min}(\rho_p,\rho_t) > 2L_t^2.
    \end{aligned}
\end{equation}
Thus, under the stated assumptions and if \eqref{eq:penalty_rules} holds, all limit points of the ADMM iterates are first-order stationary points of the smoothed problem.

The above result does not quantify how the topology of $\mathcal{G}$ affects the convergence rate.
In practice, denser connected graphs may lead to faster convergence, and related discussions can be found in~\cite{maHybridADMMUnifying2018}.

\section{NUMERICAL SIMULATIONS}
\label{sec:numerical_simulation}
\subsection{Simulation Setup}
\begin{figure}[thpb]
    \centering
    \includegraphics[width=8.4cm]{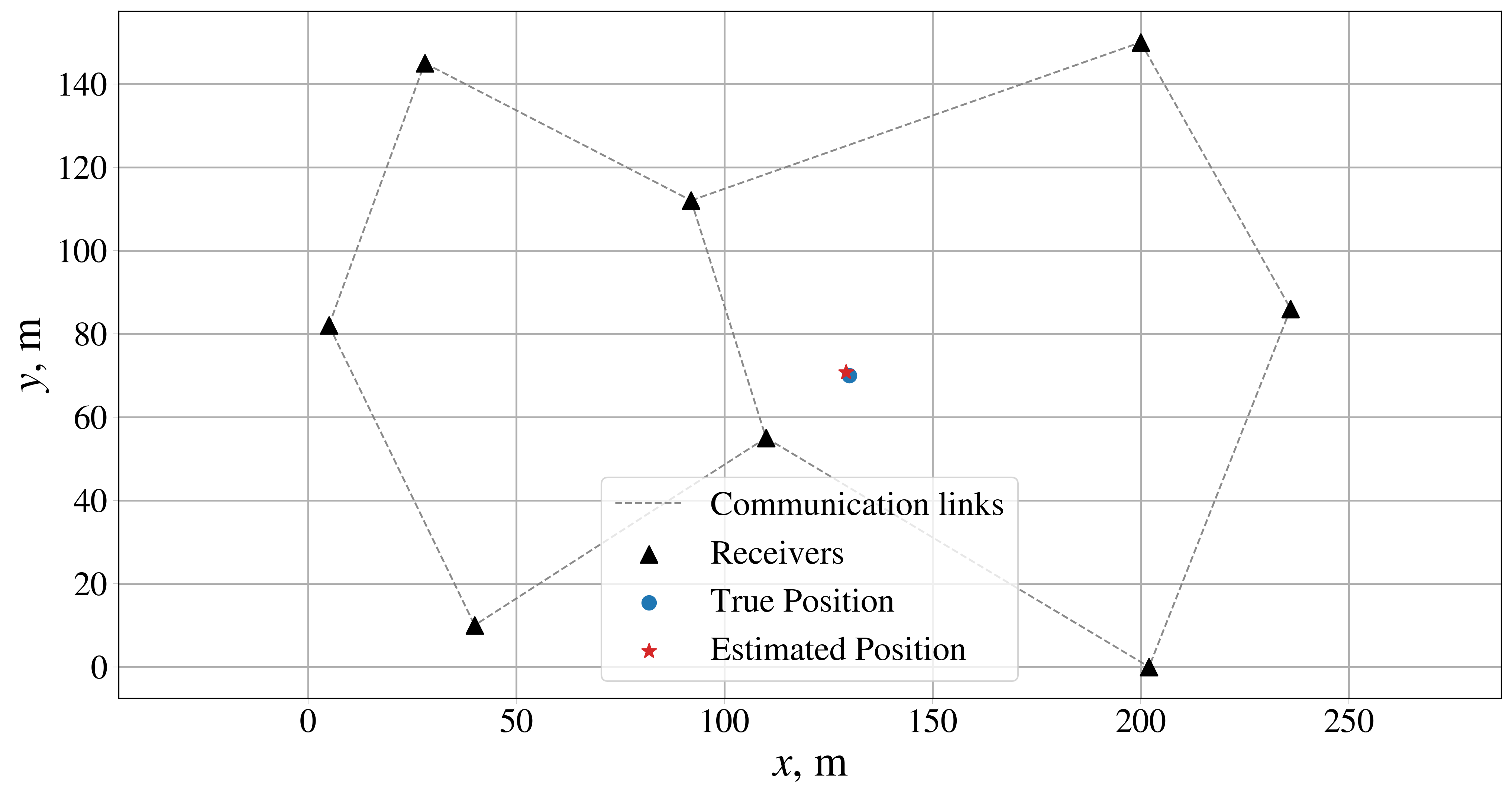}
    \caption{Receiver geometry and estimated target position in the representative single target experiment.\label{fig:oneshot_simulation}}
\end{figure}

We evaluate the proposed method in four steps: convergence for a single target realization, accuracy over random target positions, sensitivity to measurement noise relative to reference solutions and to the Cram\'er--Rao lower bound (CRLB), and sensitivity to the DADMM penalty parameters.
All simulations are conducted for a two-dimensional scene, so both the receiver positions and the target position are in $\mathbb{R}^2$.
The receiver network is fixed across all experiments and consists of eight receivers connected by a sparse communication graph, as shown in Figure~\ref{fig:oneshot_simulation}.
The sound speed is $v = 1500$~m/s, the penalty parameters are $\rho_p = 10^{-7}$ and $\rho_t = 10$, and the distributed iterations are initialized according to \eqref{eq:init_x}. 
The implementation of the proposed algorithm and the code used to conduct the experiments are available as open-source software on GitHub at \url{https://github.com/bioniwulf/playground-distributed-localization-admm}. 

\subsection{Convergence for Single Target Realization}
\begin{figure}[htbp]
  \centering
  \includegraphics[width=8.4cm]{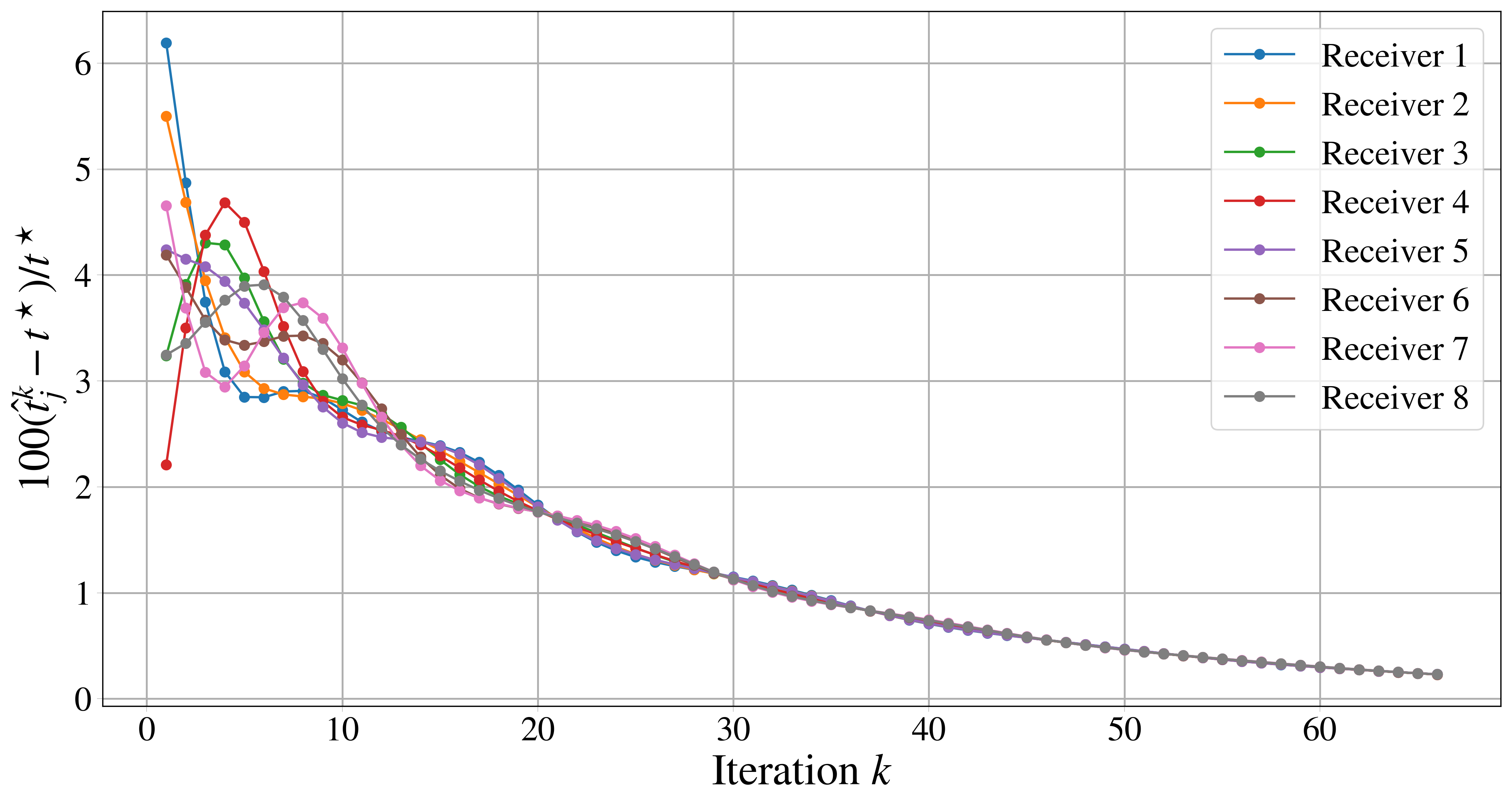}
  \medskip
  \includegraphics[width=8.4cm]{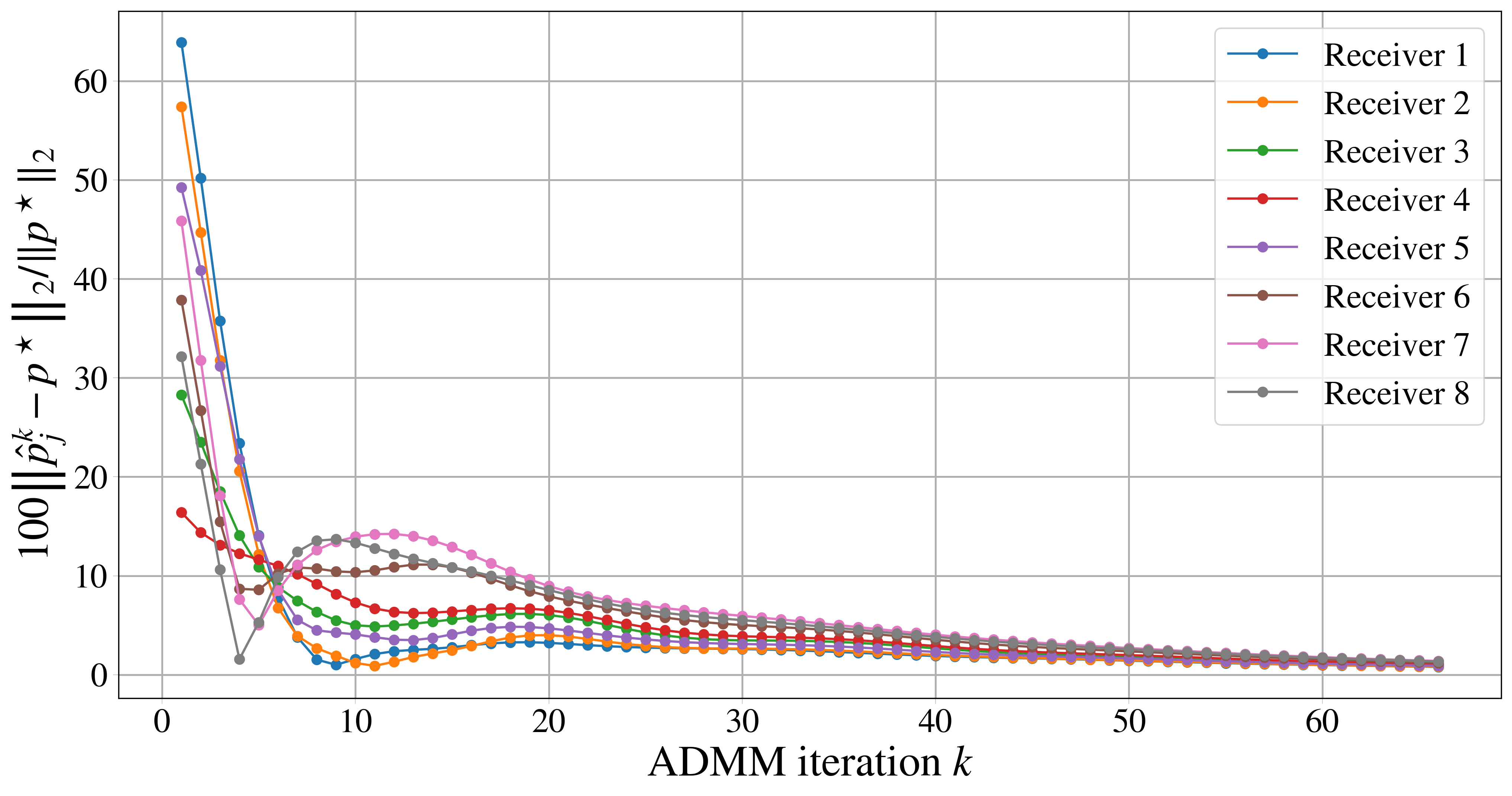}
  \caption{Evolution of the local estimates at each receiver during the one-shot experiment. Top: estimated emission time. Bottom: estimated target position.\label{fig:oneshot_convergence}}
\end{figure}

We consider a target at $\mathbf{p} = [130, 70]^\top$~m to illustrate the distributed convergence behavior.
The measurement noise standard deviation is $\sigma = 10^{-5}$~s, and the stopping thresholds are $\varepsilon_\text{feas} = \varepsilon_\text{conv} = 10^{-3}$.
Figure~\ref{fig:oneshot_convergence} shows that the node-wise estimates progressively agree and converge to a common solution.

\subsection{Accuracy over Random Target Positions}
\begin{figure}[thpb]
    \centering
    \includegraphics[width=8.4cm]{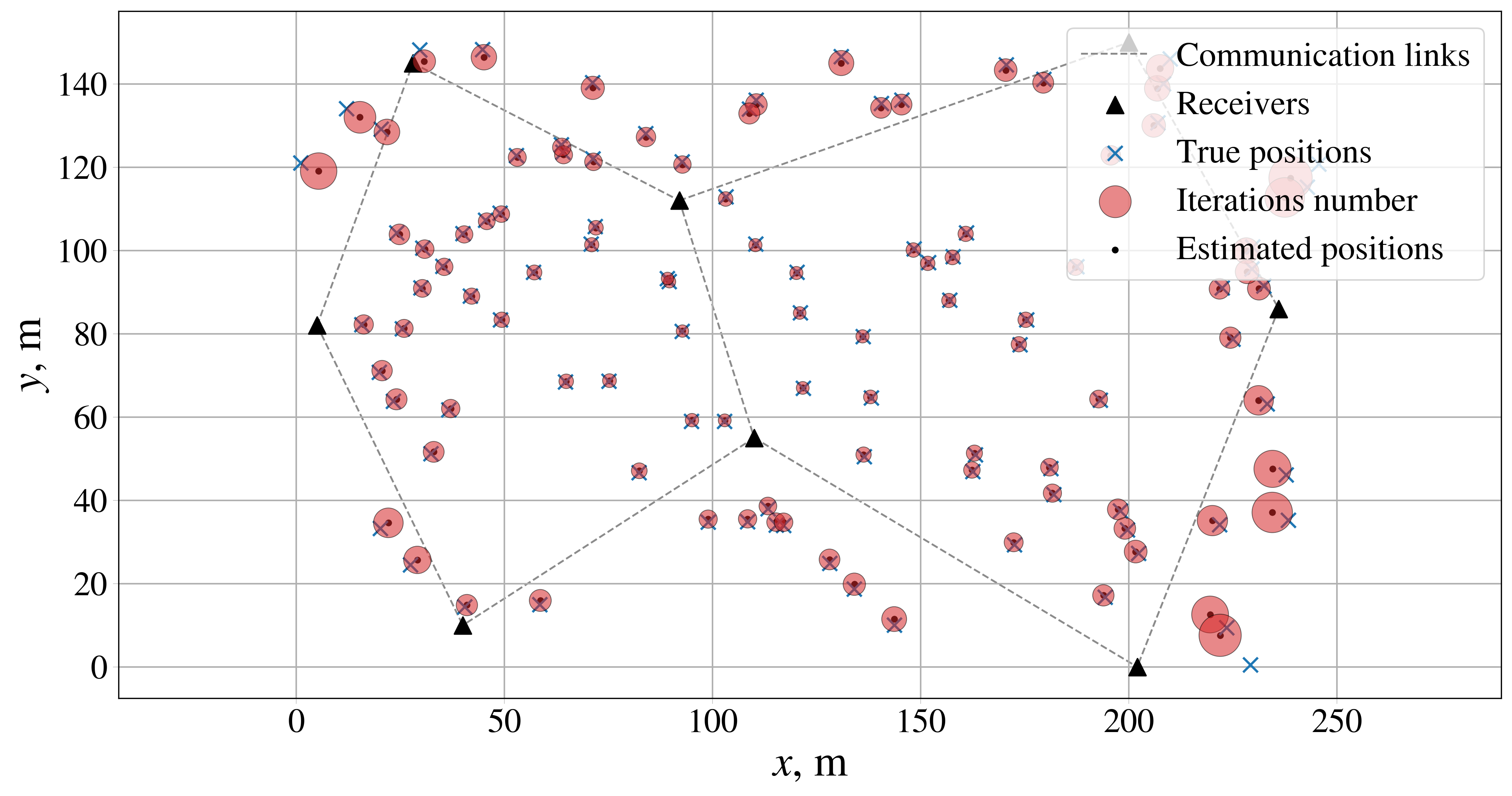}
    \caption{Localization results for randomly generated target positions in a box region around the receiver configuration. The circle radius is proportional to the number of DADMM iterations required for convergence.\label{fig:monte-carlo_field}}
\end{figure}

To assess robustness beyond a single realization, we randomly generate 100 target positions in a box surrounding the convex hull of the receivers and run DADMM from the same initialization as above.
The stopping thresholds are set to $\varepsilon_\text{feas} = \varepsilon_\text{conv} = 10^{-4}$.
Figure~\ref{fig:monte-carlo_field} shows that the estimated positions remain close to the true targets over the sampled region, while the number of iterations increases for targets farther from the receiver cluster.

\subsection{Sensitivity to Measurement Noise}
\begin{figure}[thpb]
    \centering
    \includegraphics[width=8.4cm]{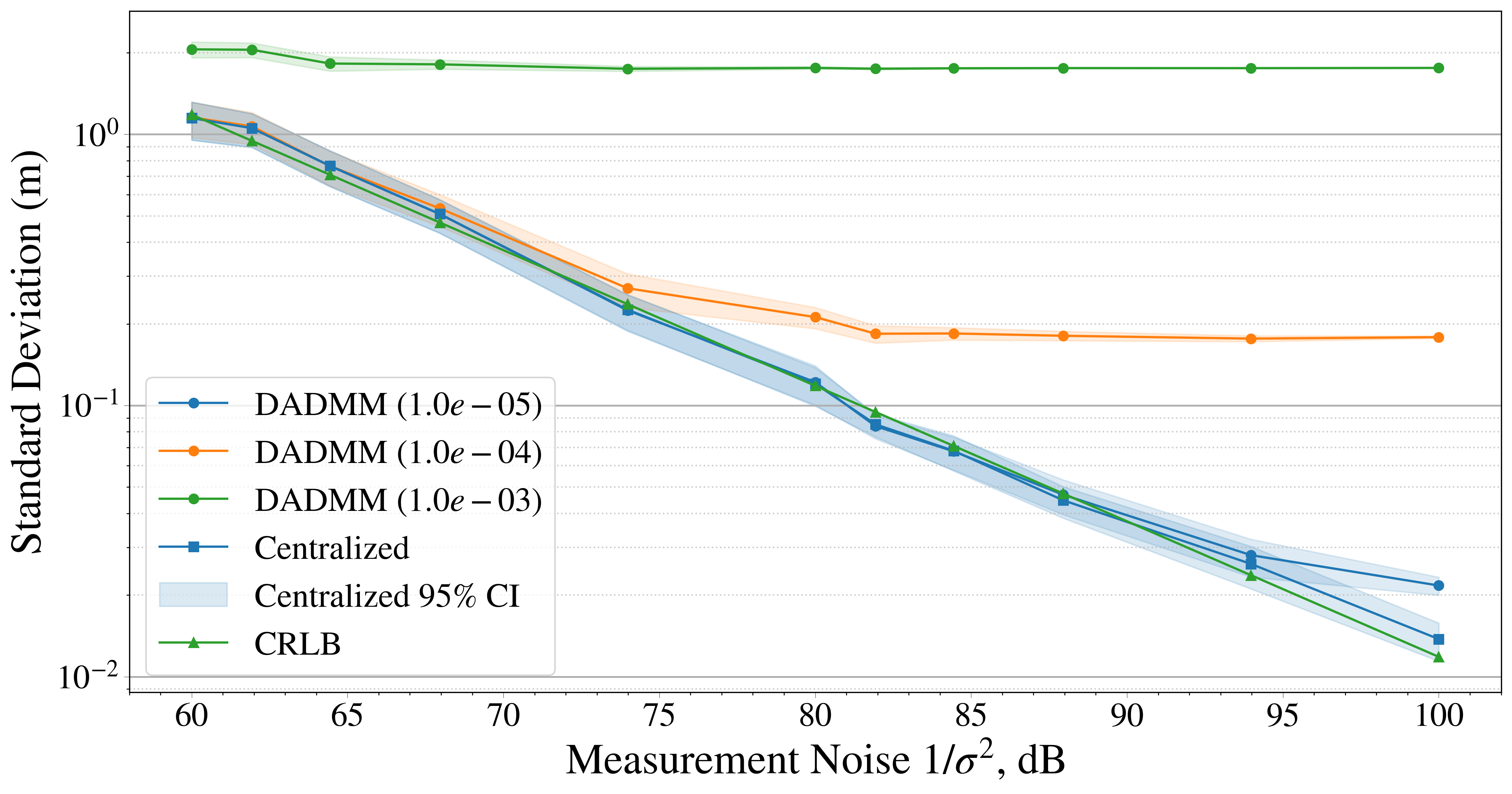}
    \caption{RMSE comparison of the proposed DADMM estimator (with three sets of stopping thresholds $\varepsilon_\text{feas} = \varepsilon_\text{conv}$), the centralized solver, and the CRLB reference under different measurement-noise levels.\label{fig:comparative}}
\end{figure}

We next compare the proposed distributed estimator against a centralized solution and against the CRLB reported in~\cite{xuSourceLocalizationWireless2011} for a closely related centralized ToA formulation with unknown time of signal emission.
The centralized estimate is obtained by solving the nonconvex problem~\eqref{eq:opt_problem} with IPOPT \cite{wachterImplementationInteriorpointFilter2006a} through CasADi \cite{anderssonCasADiSoftwareFramework2019a}, using the centroid of the receiver network and the average signal-arrival time for initialization.
The measurement-noise standard deviation varies from $10^{-5}$~s to $10^{-3}$~s. For the distributed method, the stopping thresholds are set to $\varepsilon_\text{feas} = \varepsilon_\text{conv} \in \{10^{-3}, 10^{-4}, 10^{-5}\}$.
Figure~\ref{fig:comparative} shows that the distributed RMSE follows the same overall trend as the centralized benchmark and the CRLB reference, while tighter stopping thresholds improve accuracy at the cost of more DADMM iterations.

\subsection{Penalty Parameters Analysis}
\begin{figure}[htbp]
  \centering
  \includegraphics[width=8.4cm]{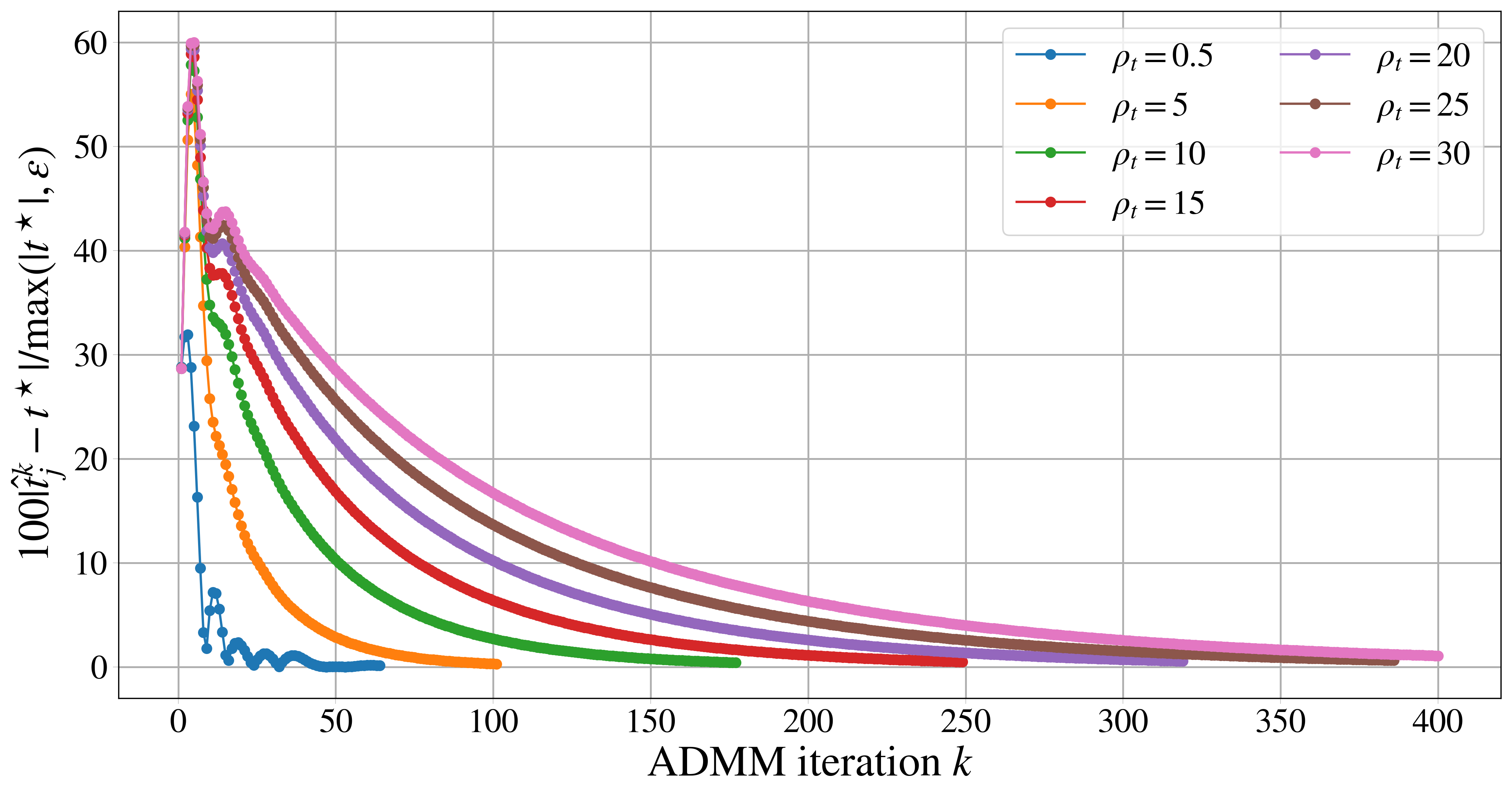}
  \medskip
  \includegraphics[width=8.4cm]{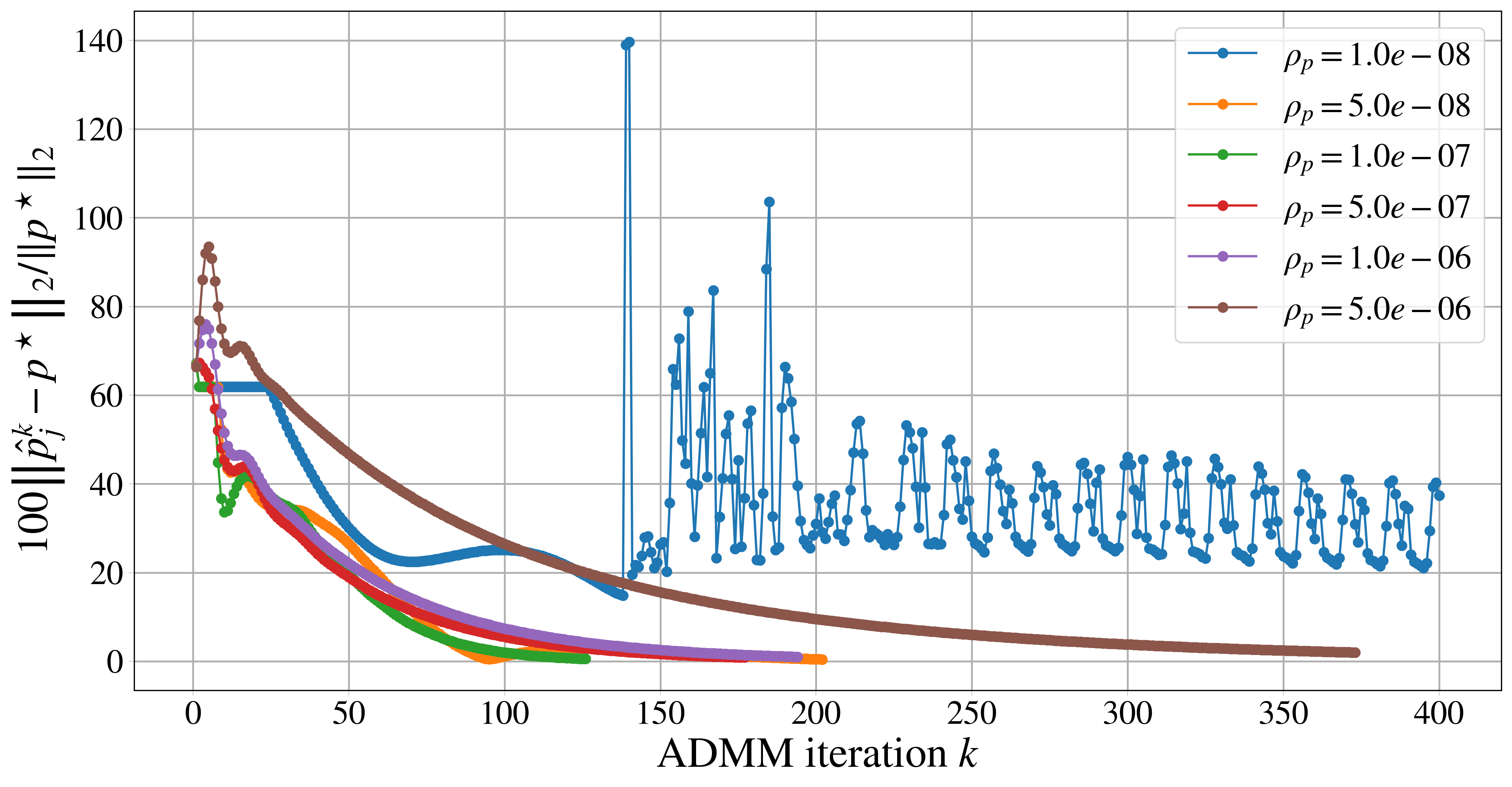}
  \caption{Evolution of the local estimates at a representative receiver for different values of the penalty parameters. Top: estimated emission time evolution with $\rho_p = 10^{-7}$. Bottom: estimated target position evolution with $\rho_t = 10$.\label{fig:penalty_evolution}}
\end{figure}

Finally, we examine the influence of $\rho_p$ and $\rho_t$ on the convergence behavior.
Figure~\ref{fig:penalty_evolution} shows the emission-time convergence for varying $\rho_t$ with fixed $\rho_p = 10^{-7}$, and the position convergence for varying $\rho_p$ with fixed $\rho_t = 10$.
Small penalties lead to oscillatory or weakly damped behavior and may prevent consensus within the iteration budget, whereas overly large penalties slow the convergence.
This empirical behavior is qualitatively consistent with the role of the penalty parameters in the smoothed-problem analysis of \eqref{eq:penalty_rules}, although the theoretical bounds there may be conservative in practice.

\section{CONCLUSIONS AND FUTURE WORK}
\label{sec:conclusion}
This paper addressed the problem of distributed localization of a non-cooperative acoustic source from synchronized ToA measurements with unknown time of signal emission.
By reformulating the problem directly in the ToA domain and jointly estimating the emission time, the proposed approach avoids the structural mismatch induced by pairwise TDoA preprocessing and enables a distributed DADMM solution over the receiver network.
The resulting algorithm admits closed-form local update equations without nested numerical subproblems.
We established convergence properties and sufficient conditions on the penalty parameters for a smoothed approximation of the measurement model.

The numerical results show that the local estimates converge to a common consensus solution, remain accurate over a broad range of target positions, and achieve performance comparable to that of a centralized solver while following the same overall trend as the considered CRLB reference.
They also illustrate the practical effect of stopping thresholds and penalty parameters on the trade-off between accuracy, convergence speed, and numerical stability.

A natural direction for future work is to extend the same distributed optimization framework to optimization-based tracking methods, in particular Moving Horizon Estimation, whose problem structure is closely related to the formulation considered here.

\section{ACKNOWLEDGMENTS}
\label{sec:acknowledgments}
The authors would like to thank Professor João M. F. Xavier (Instituto Superior Técnico, University of Lisbon, Portugal) for his valuable guidance and insightful advice during this research.

\appendix
\subsection{Proximal step for \texorpdfstring{$\mathbf{x}^{[k+1]}$}{x[k+1]}}
\label{sec:app_proximal_x}
Consider the optimization problem in~\eqref{eq:prox_x}, where the index $i$ is omitted for clarity of exposition, that is,
\begin{equation}
    \label{eq:appx_prox_x}
    \mathbf{x}^{[k+1]} = \arg \min_{\mathbf{x}}
        f(\mathbf{x}) + \frac{1}{2}\sum_{j \in \mathcal{N}}
        \|\mathbf{x} - \mathbf{y}_j^{[k]} + \mathbf{u}_{j}^{[k]}\|_\Theta^2
\end{equation}
with
\[
f(\mathbf{x}) = \frac{1}{2} \left( \tau - t - \tfrac{1}{v}\|\mathbf{p} - \mathbf{s}\| \right)^2.
\]
Let $\mathbf{a}_j = \mathbf{y}_j^{[k]} - \mathbf{u}_{j}^{[k]}$ and define the average vector
\[
    \bar{\mathbf{a}} =
    \begin{bmatrix}
        \bar{\mathbf{a}}_p \\
        \bar{a}_t
    \end{bmatrix}
    =
    \frac{1}{|\mathcal{N}|} \sum_{j \in \mathcal{N}} \mathbf{a}_j
    =
    \frac{1}{|\mathcal{N}|} \sum_{j \in \mathcal{N}} (\mathbf{y}_j^{[k]} - \mathbf{u}_{j}^{[k]}).
\]
Then, the quadratic penalty term can be rewritten as
\begin{equation}
    \label{eq:appdx_neighbor_sum}
    \frac{1}{2}\sum_{j \in \mathcal{N}}
    \|\mathbf{x} - \mathbf{y}_j^{[k]} + \mathbf{u}_{j}^{[k]}\|_\Theta^2
    =
    \frac{|\mathcal{N}|}{2} \|\mathbf{x} - \bar{\mathbf{a}}\|_\Theta^2
    +
    C,
\end{equation}
for a constant $C$ that does not depend on $\mathbf{x}$.

Next, we reparameterize the target position as
$
\mathbf{p} = \mathbf{s} + r\mathbf{n},
$
where $r = \|\mathbf{p} - \mathbf{s}\| \ge 0$ and $\|\mathbf{n}\| = 1$.
By discarding the constant $C$ and introducing $\mathbf{d} = \bar{\mathbf{a}}_p - \mathbf{s}$,~\eqref{eq:appx_prox_x} becomes
\begin{equation}
    \label{eq:appx_prox_x_ref}
    \begin{aligned}
        \min_{r, \mathbf{n}, t}
        \quad &
        \frac{1}{2} \left( \tau - t - \tfrac{r}{v} \right)^2
        + \frac{\rho_p |\mathcal{N}|}{2}\|r\mathbf{n} - \mathbf{d}\|^2
        + \frac{\rho_t |\mathcal{N}|}{2}(t - \bar{a}_t)^2 \\
        \text{s.t.} \quad &
        \|\mathbf{n}\| = 1,\ r \ge 0.
    \end{aligned}
\end{equation}
For fixed $r$ and $t$, the only term that depends on $\mathbf{n}$ is $\|r\mathbf{n} - \mathbf{d}\|^2$.
Since
\[
\|r\mathbf{n} - \mathbf{d}\|^2 = r^2 + \|\mathbf{d}\|^2 - 2r\,\mathbf{d}^\top \mathbf{n},
\]
its minimum over the unit sphere is attained when $\mathbf{n}$ is aligned with $\mathbf{d}$.
Hence, for $\mathbf{d} \neq 0$,
\begin{equation}
    \label{eq:appx_n_solution}
    \mathbf{n}^\star = \frac{\mathbf{d}}{\|\mathbf{d}\|}.
\end{equation}
Substituting~\eqref{eq:appx_n_solution} into~\eqref{eq:appx_prox_x_ref} yields the reduced problem
\begin{equation}
    \label{eq:appx_prox_x_ref_simple}
    \begin{aligned}
        \min_{r, t}
        \quad &
        \frac{1}{2} \left( \tau - t - \tfrac{r}{v} \right)^2
        + \frac{\rho_p |\mathcal{N}|}{2}(r - \|\mathbf{d}\|)^2
        + \frac{\rho_t |\mathcal{N}|}{2}(t - \bar{a}_t)^2 \\
        \text{s.t.} \quad &
        r \ge 0.
    \end{aligned}
\end{equation}
This is a strictly convex quadratic optimization problem in $(t,r)$.
Its unconstrained minimizer is obtained from the first-order optimality conditions, which lead to the linear system
\begin{equation}
    \label{eq:linear_solution}
    \begin{bmatrix}
        1 + \rho_t |\mathcal{N}| & 1/v \\
        1/v & 1/v^2 + \rho_p |\mathcal{N}|
    \end{bmatrix}
    \begin{bmatrix}
        t^* \\
        r^*
    \end{bmatrix}
    =
    \begin{bmatrix}
        \tau + \rho_t |\mathcal{N}| \bar{a}_t \\
        \tau/v + \rho_p |\mathcal{N}| \|\mathbf{d}\|
    \end{bmatrix}.
\end{equation}
If $r^* \ge 0$, which is the nominal case in the localization setting considered here, then $(t^*, r^*)$ is also the unique minimizer of~\eqref{eq:appx_prox_x_ref_simple}; otherwise, the constrained minimizer is attained at the boundary $r = 0$.

Therefore, for $\mathbf{d} \neq 0$ and $r^* \ge 0$, the solution of~\eqref{eq:appx_prox_x} is
\begin{equation}
    \label{eq:appx_prox_x_solution}
    \mathbf{x}^{[k+1]}
    =
    \begin{bmatrix}
        \mathbf{s} + r^* \dfrac{\mathbf{d}}{\|\mathbf{d}\|} \\
        t^*
    \end{bmatrix},
\end{equation}
with $(t^*, r^*)$ obtained from~\eqref{eq:linear_solution}.
Restoring the node index $i$ gives the update formulas used in Section~\ref{sec:localization_decentralized}.

This expression is well-defined for $\mathbf{d} \neq 0$, which is ensured by the initialization in~\eqref{eq:init_x} except for the degenerate centroid case discussed in Section~\ref{sec:localization_decentralized}.

\subsection{Lipschitz Constants for the ToA Residual Function}
\label{sec:app_lipschitz}

A twice continuously differentiable function $f:\mathbb{R}^n \to \mathbb{R}$ is said to be Lipschitz-differentiable on an analysis region $\Omega$ if there exists a constant $L>0$ such that
\begin{equation}
    \|\nabla^2 f(x)\| \le L, 
    \quad \forall x \in \Omega,
\end{equation}
where $\|\cdot\|$ denotes the spectral norm.

Consider the smoothed objective function for a single agent defined in~\eqref{eq:optimization_function_smoothed}.
We omit the index $i$ to simplify the presentation and write
\begin{equation}
    \label{eq:lfunction}
    f(\mathbf{p},t)
    =
    \frac{1}{2}
    (
    \tau - t - \frac{1}{v} r(\mathbf{p})
    )^2,
\end{equation}
where $r(\mathbf{p}) = \sqrt{\|\mathbf{p} - \mathbf{s}\|^2 + \varepsilon}$.

The Hessian of the function $f(\mathbf{p},t)$ has the block structure
\begin{equation}
\label{eq:nabla_f}
\nabla^2 f(\mathbf{p},t)
=
\begin{bmatrix}
\nabla^2_{pp} f & \nabla^2_{pt} f \\
\nabla^2_{tp} f & \nabla^2_{tt} f
\end{bmatrix}.
\end{equation}

The second derivative with respect to time is obtained directly from~\eqref{eq:lfunction} as
\begin{equation}
\label{eq:nabla_f_t}
\nabla^2_{tt} f(\mathbf{p},t)
=
1.
\end{equation}

The gradient with respect to $\mathbf{p}$ is
\begin{equation}
\label{eq:f_nabla_p}
\nabla_{\mathbf{p}} f(\mathbf{p},t)
=
-
\frac{1}{v}
(
\tau - t - \frac{1}{v} r(\mathbf{p})
)
\nabla_{\mathbf{p}} r(\mathbf{p}),
\end{equation}
where
\begin{equation}
\nabla_{\mathbf{p}} r(\mathbf{p})
=
\frac{\mathbf{p}-\mathbf{s}}{r(\mathbf{p})}.
\end{equation}

Differentiating once more with respect to $\mathbf{p}$ yields
\begin{equation}
\label{eq:nabla_f_pp}
\nabla^2_{pp} f(\mathbf{p},t)
=
\frac{1}{v^2 }
\nabla_{\mathbf{p}} r(\mathbf{p})
\nabla_{\mathbf{p}} r(\mathbf{p})^\top
-
\frac{e}{v}
\nabla^2_{pp} r(\mathbf{p}),
\end{equation}
where
\begin{equation}
e = \tau - t - \frac{1}{v} r(\mathbf{p}),
\end{equation}
and
\begin{equation}
\nabla^2_{pp} r(\mathbf{p})
=
\frac{1}{r(\mathbf{p})} I
-
\frac{1}{r(\mathbf{p})^3}
(\mathbf{p}-\mathbf{s})(\mathbf{p}-\mathbf{s})^\top.
\end{equation}

The mixed derivatives follow from~\eqref{eq:f_nabla_p}:
\begin{equation}
\nabla^2_{tp} f(\mathbf{p},t)
=
\nabla^2_{pt} f(\mathbf{p},t)
=
\frac{1}{v}
\nabla_{\mathbf{p}} r(\mathbf{p}).
\end{equation}

In accordance with~\eqref{eq:r_relaxation}, we assume $r(\mathbf{p}) \ge \sqrt{\varepsilon}$.
Hence,
\begin{equation}
\label{eq:r_upperbound}
\|\nabla_{\mathbf{p}} r(\mathbf{p})\| \le 1,
\quad
\|\nabla^2_{pp} r(\mathbf{p})\|
\le
\frac{1}{r(\mathbf{p})}
\le
\frac{1}{\sqrt{\varepsilon}}.
\end{equation}

\subsubsection*{Lipschitz constant with respect to \texorpdfstring{$t$}{t}}
Using~\eqref{eq:nabla_f} and~\eqref{eq:nabla_f_t}, we obtain
\begin{equation}
\|\nabla^2_{tt} f(\mathbf{p}, t)\|
\le
1
=: L_t.
\end{equation}

\subsubsection*{Lipschitz constant with respect to \texorpdfstring{$\mathbf{p}$}{p}}

From~\eqref{eq:nabla_f_pp} we have
\begin{equation}
\label{eq:nabla_pp_Fnitial}
\|\nabla^2_{pp} f(\mathbf{p}, t)\|
\le
\left(
\frac{1}{v^2}
\|\nabla_{\mathbf{p}} r(\mathbf{p})\|^2
+
\frac{|e|}{v}
\|\nabla^2_{pp} r(\mathbf{p})\|
\right).
\end{equation}

We impose the trust-region condition
\begin{equation}
\label{eq:e_trust}
|e| \le k \sigma,
\end{equation}
where $k$ is a confidence multiplier scaled by the measurement-noise standard deviation $\sigma$.

Substituting~\eqref{eq:r_upperbound}, \eqref{eq:e_meaning}, and~\eqref{eq:e_trust} into~\eqref{eq:nabla_pp_Fnitial}, we obtain
\begin{equation}
\|\nabla^2_{pp} f(\mathbf{p}, t)\|
\le
\left(
\frac{1}{v^2}
+
\frac{k\sigma}{v R_\text{min}}
\right)
=: L_p.
\end{equation}

Therefore, on the analysis region $\Omega$, the function $f(\mathbf{p},t)$ is Lipschitz-differentiable with block-wise constants $L_t$ and $L_p$.

\bibliographystyle{IEEEtran}
\bibliography{references}

\end{document}